\documentclass[prl,twocolumn,showpacs]{revtex4}
\def\cuscn{$\kappa$-(BEDT-TTF)$_2$Cu(NCS)$_2$}
\newcommand{\gtsim}{\mbox{{\raisebox{-0.4ex}{$\stackrel{>}{{\scriptstyle\sim}}
$}}}}
\newcommand{\ltsim}{\mbox{{\raisebox{-0.4ex}{$\stackrel{<}{{\scriptstyle\sim}}
$}}}}
\usepackage{graphicx}
\usepackage{dcolumn}
\usepackage{amsmath}
\usepackage{longtable}
\begin{document}

\title{Angle-dependent 
magnetoresistance oscillations due to magnetic breakdown orbits}
\author{A. F. Bangura$^{1,2}$, P. A. Goddard$^{1}$, 
J. Singleton$^{3}$, S. W. Tozer$^{4}$, A.~I.~Coldea$^{1,2}$, 
A. Ardavan$^{1}$, R.D. McDonald$^{3}$, S. J. Blundell$^1$ and J.A. Schlueter$^5$}
\affiliation{$^{1}$Clarendon Laboratory, University of Oxford, 
Parks Road, Oxford OX1~3PU, UK\\
$^{2}$H. H. Wills Physics Laboratory, University of Bristol, 
Tyndall Avenue, BS8 1TL, UK\\
$^{3}$National High Magnetic Field Laboratory, 
Los Alamos National Laboratory, TA-35, MS-E536, Los Alamos, NM~87545 USA\\
$^{4}$National High Magnetic Field Laboratory, 
Tallahassee, FL 83810 USA\\
$^5$Materials Science Division, Argonne National Laboratory,
Argonne, Illinois 60439, USA
}

\begin{abstract}
We present experimental evidence for a 
hitherto unconfirmed type of 
angle-dependent magnetoresistance oscillation
caused by magnetic breakdown.
The effect was observed in the
organic superconductor \cuscn~
using hydrostatic pressures of up to 9.8 kbar
and magnetic fields of up to 33~T.
In addition, we show that similar oscillations are 
revealed in ambient pressure measurements, provided that the 
Shubnikov-de Haas oscillations are suppressed either by 
elevated temperatures or filtering of the data. 
These results provide a compelling validation of 
Pippard's semiclassical picture of magnetic breakdown.
\end{abstract}
\pacs{71.18.+y, 71.20.Rv, 72.15.Gd, 74.25.Jb}

\maketitle
Recently, the measurement of angle-dependent 
magnetoresistance oscillations (AMROs)
has emerged as a powerful technique in the elucidation of 
the fine details of Fermi surfaces in reduced-dimensionality 
metals~\cite{osada,ruthenates,Kartsovnik2004}.
In contrast to de Haas-van Alphen oscillations,
AMROs can be observed in rather low-quality 
samples~\cite{Hussey2003,Blundell1996}, or
when temperatures are relatively high~\cite{Goddard2004}; 
they have therefore
been measured in a wide variety of systems,
including crystalline organic metals~\cite{Kartsovnik2004,Singleton2000},
ruthenates~\cite{ruthenates},
semiconductor superlattices~\cite{osada}, and
cuprate superconductors~\cite{Hussey2003}.
AMROs can, in most cases~\cite{Blundell1996}, be attributed to
the time-evolution of the quasiparticle velocities
as they traverse the Fermi surface under 
the influence of the magnetic field. 
Consequently, and based on the topologies of the orbits involved,
several distinct species of AMRO have been identified,
including Lebed~\cite{Lebed1989,LMAcomment} 
and Danner-Chaikin-Kang (DCK) oscillations~\cite{Danner1994} 
due to orbits on quasi-one-dimensional (Q1D) sections of
Fermi surface, and Yamaji oscillations~\cite{Kartsovnik1988,Yamaji1989,Yagi1991},
associated with closed orbits on quasi-two-dimensional (Q2D) 
Fermi-surface sections (for a detailed description of the differences between these 
effects see~\cite{Kartsovnik2004}).
In this Letter we report the measurement of a further class
of AMROs, observed only at high magnetic fields
and caused by magnetic breakdown.

\begin{figure}[t]
\centering
\includegraphics[width=3.3cm]{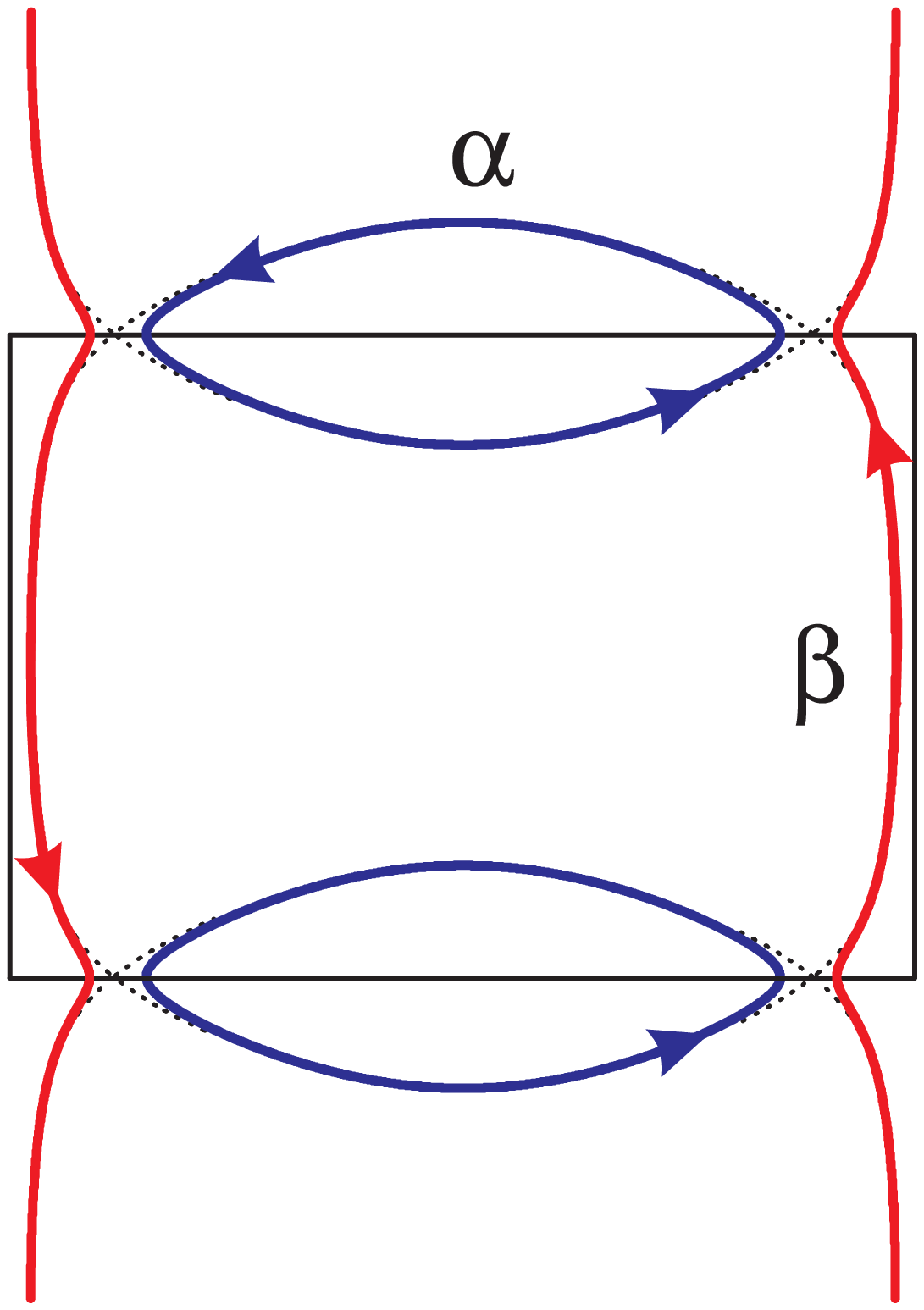}
\includegraphics[width=5.2cm]{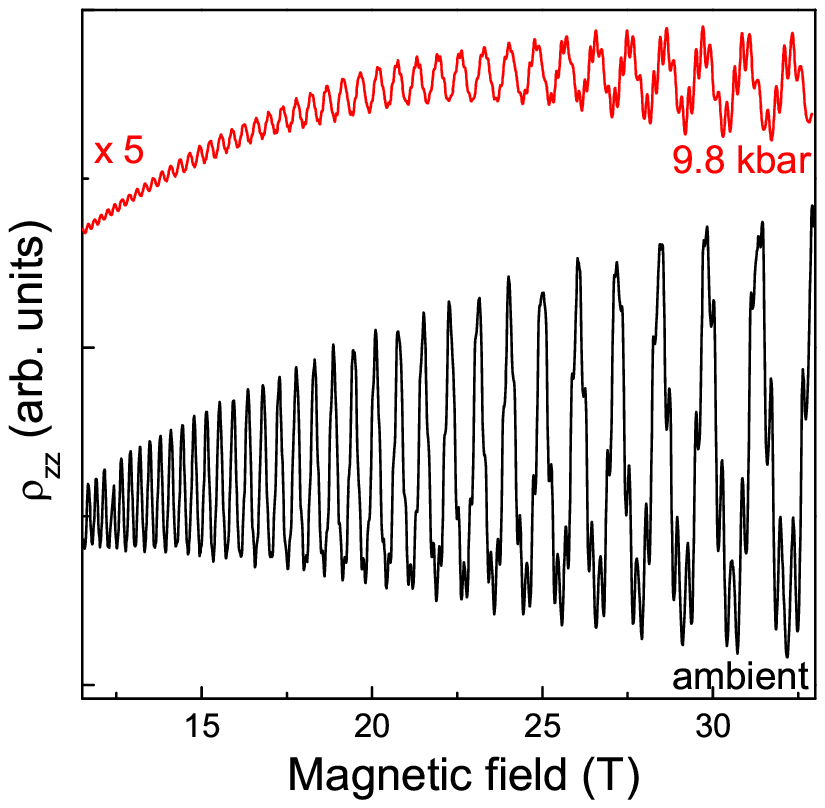}
\caption{Left: (Color online) Fermi surface cross-section
of \cuscn~ in the Q2D planes, showing the Q2D pockets (blue), 
the Q1D sheets (red), the Brillouin zone boundaries (black)
and the $\beta$ breakdown and $\alpha$ orbits. 
Right: Shubnikov-de Haas oscillations observed 
in two \cuscn~samples at $T\approx500$~mK, 
with the magnetic field directed perpendicular to the 
conducting planes ($\theta=0^\circ$). 
The upper trace is for a sample at a
pressure $p=9.8 \pm 0.2$~kbar;
the other is at ambient pressure.  
At low fields a single oscillation frequency is 
present corresponding to the $\alpha$-orbit. 
At larger fields, higher frequencies 
are seen, corresponding to the $\beta$-orbit 
and other magnetic breakdown orbits~\cite{Harrison1996}. 
The 9.8~kbar data are enhanced by a factor of 5 
and the curves are offset for clarity.}
 \label{fig1}
\vspace{-7mm}
\end{figure}

The crystalline organic metal \cuscn~was 
chosen for the experiments
because its Fermi surface both
resembles the coupled network model for magnetic breakdown
first proposed by Pippard~\cite{Pippard1962}
and is very well characterized by 
theory~\cite{Oshima1988} and 
experiment~\cite{Harrison1996,Toyota1989,Caulfield1994,Goddard2004}.
The Fermi surface is shown in Fig.~\ref{fig1}; it comprises
a Q2D pocket and a pair of Q1D sheets. The Q2D and Q1D sections 
are separated in $k$-space at the Brillouin-zone 
boundary owing to a weak periodic potential caused by
the translational symmetry of the anion layers~\cite{Singleton2000}.
At sufficiently high magnetic fields $B$, 
mixing between the states
on the two Fermi surface sections leads to 
magnetic breakdown, in which a quasiparticle
``tunnels'' in $k$-space between 
them~\cite{Pippard1962,Harrison1996,Shoenberg1984}.
In Pippard's semiclassical picture~\cite{Pippard1962},
this enables quasiparticles to execute 
the large $\beta$ orbit (Fig.~\ref{fig1})
and other more complex orbits about the Fermi surface,
leading to the observation of high-frequency Shubnikov-de Haas
(SdH) and de Haas-van Alphen 
oscillations~\cite{Harrison1996,Kartsovnik1996,Mihut2006}.
The probability
\begin{equation}
P=\exp(-B_0/B)
\label{bd}
\end{equation}
of magnetic breakdown
is parameterized by $B_0$, the characteristic 
{\it breakdown field}~\cite{Pippard1962,Harrison1996,Shoenberg1984}.
In this Letter, we show that magnetic breakdown can additionally produce a new type of AMRO in \cuscn. The origin of this phenomenon is
similar to that of Yamaji oscillations~\cite{Goddard2004}
but in the present case, the quasiparticle trajectories responsible
are {\it magnetic breakdown orbits}, rather than closed paths on Q2D
Fermi-surface sections. In order to distinguish the new features
from the more conventional Lebed or Yamaji oscillations 
we will refer to them as breakdown-AMROs or BAMROs.

Four-wire magnetotransport experiments are 
performed on single crystals of \cuscn~in 
quasistatic fields produced by 33~T Bitter 
coils and the 45~T hybrid magnet at NHMFL Tallahassee. 
A two-axis goniometer allows continuous rotation 
of the angle $\theta$ between the applied magnetic field 
and the normal to the highly conductive planes 
of the sample, as well as discrete changes in the 
plane of rotation parameterized by the azimuthal angle $\phi$. 
(In \cuscn~ we define the $\phi=0^\circ$ plane of rotation 
as being perpendicular to the Q1D sheets.) 
The goniometer is placed within a $^3$He cryostat allowing 
temperatures $T$ as low as 500 mK. 
Electrical contacts are applied to the samples using $12.5~\mu$m Au or 
Pt wires attached using graphite paint. 
For the high pressure measurements, the samples are placed 
inside a miniature anvil cell of length 9~mm and outer 
diameter 6~mm~\cite{Tozer1993}. Pressure ($p$)
measurement is performed 
{\it in-situ} using the ruby fluorescence line at $\approx$ 690~nm, 
excited using the 448~nm line of an Ar-ion laser;
the pressure dependence of this ruby line is well known~\cite{Barnett1973}. 
A single optical fiber is used to excite and collect the 
fluorescence of a chip of ruby placed within the cell 
next to the sample, and is compared to that of a 
chip at the same $T$ 
outside the pressure cell. 
Typical \cuscn~sample dimensions are 
$\sim 0.7\times 0.5\times 0.1$~mm$^3$ for the ambient-pressure
experiments, and $\sim 0.1\times 0.1\times 0.04$~mm$^3$ 
for the high-pressure measurements.

Data such as those in Fig.~\ref{fig1} were Fourier-analysed
to reveal the SdH oscillation frequencies $F$
present. In addition to frequencies due to the classically-allowed
$\alpha$ orbit about the Q2D pocket ($F_{\alpha}$), and the $\beta$ breakdown
orbit ($F_{\beta}$), combination frequencies such as 
$F_{\beta}-F_{\alpha}$ and $F_{\beta}-2F_{\alpha}$
caused by the Shiba-Fukuyama-Stark quantum interference 
effect~\cite{Singleton2000,Harrison1996,Kartsovnik1996,Mihut2006} 
are observed in the
Fourier transforms.
The frequencies found were $F_\alpha = 750 \pm 20$~T
and $F_\beta = 4030 \pm 60$~T at $p=9.8$~kbar,
and $F_\alpha = 610 \pm 10$~T and $F_\beta = 3950 \pm 30$~T
at ambient pressure.
In addition, the $B$ and $T$ dependences
of the $F_{\alpha}$ frequency Fourier amplitudes
were fitted using the standard Lifshitz-Kosevich 
formalism appropriate for a 2D metal~\cite{Harrison1996,Shoenberg1984}. 
In this way the effective mass $m^*_\alpha$  at $\theta=0^\circ$ 
and ``Dingle'' scattering time $\tau_\alpha$~\cite{Shoenberg1984,SingletonEncy}
of the $\alpha$-pocket quasiparticles 
at 9.8~kbar were determined to be $2.0 \pm 0.1~m_{\rm e}$ and 
$0.81 \pm 0.05$~ps, respectively, where $m_{\rm e}$ is the 
electron rest mass; equivalent values for the
ambient-pressure experiments were
$3.5 \pm 0.1~m_{\rm e}$ and $2.3 \pm 0.2$~ps. 
These masses and frequencies 
are in reasonable agreement with previous high-pressure SdH 
data~\cite{Caulfield1994}. 

\begin{figure}[t]
\centering
   \includegraphics[height=5cm]{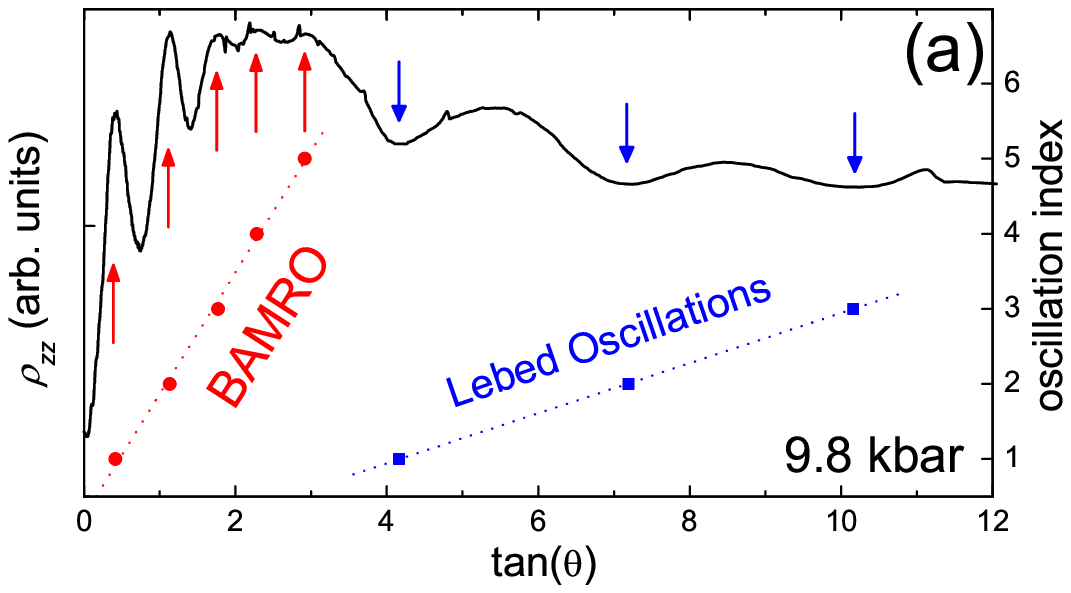}
   \includegraphics[height=6cm]{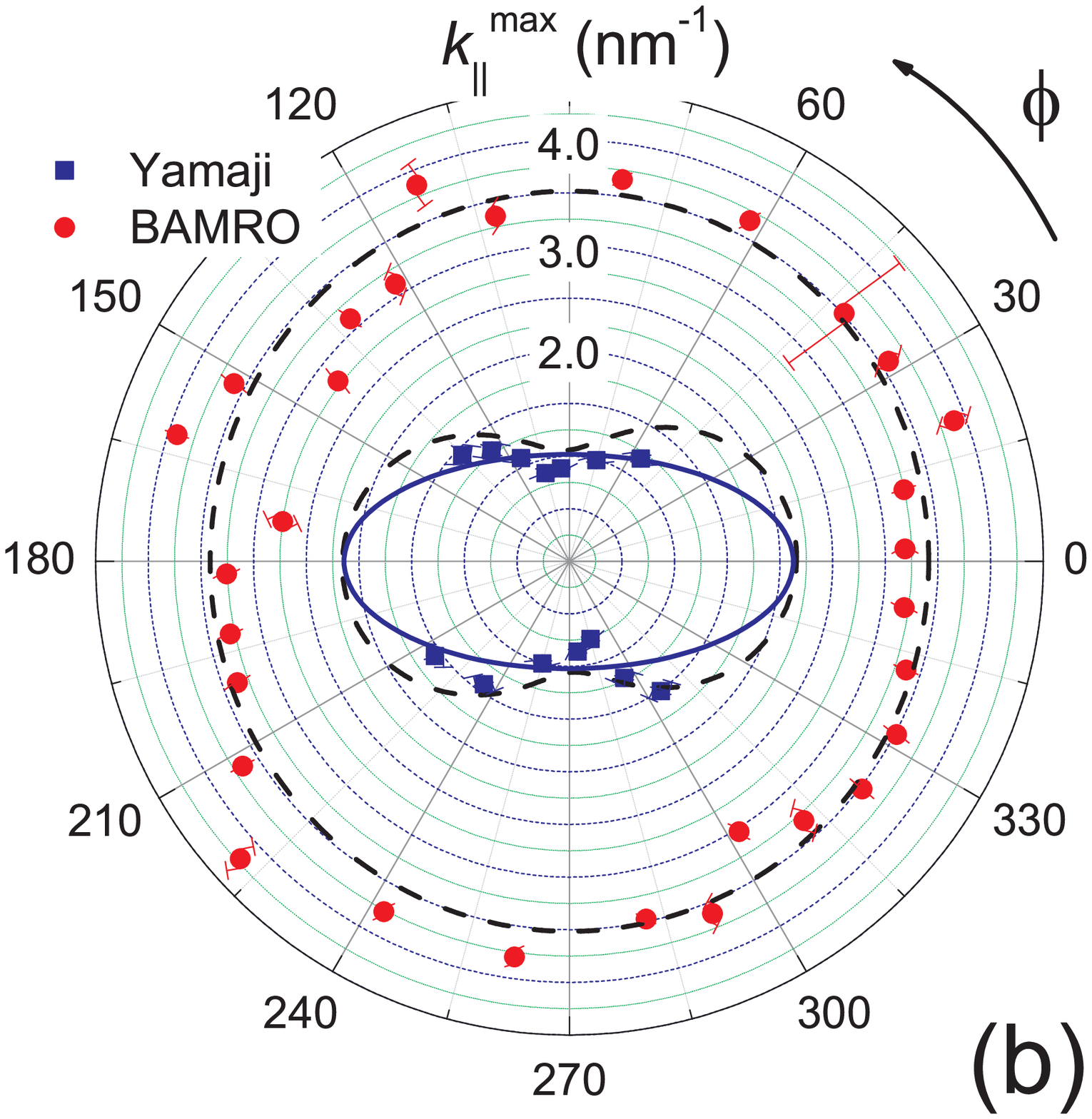}
\vspace{-5mm}
\caption{(Color online) (a) Magnetoresistance of \cuscn~(black curve) as a 
function of $\tan\theta$ at 
$p=9.8$~kbar, $B=30$~T, $T=1.5$~K  and $\phi=160^\circ$.  
The positions of the 
BAMROs are marked with up (red) arrows and the Lebed
oscillations with down (blue) arrows. 
Also shown are these positions as a function of 
oscillation index (points) from which the 
frequencies may be extracted. (b) Polar plot of the 
maximum in-plane Fermi wavevector, $k^{\rm max}_{||}$, 
derived from the $\tan\theta$ oscillation frequency, 
as a function of $\phi$. Blue squares are Yamaji oscillations 
and red circles are BAMROs. Fits of Eqn.~\ref{kmax} to the 
data (dashed lines) allow the geometry of the orbits which 
give rise to the oscillations to be determined. 
The dimensions of the $\alpha$-orbit thus derived are shown (blue line). 
The error in $\phi$ is $\pm$ 5$^{\circ}$.}
\label{fig2}
\vspace{-7mm}
\end{figure}

An earlier study of AMROs in ambient-pressure \cuscn~\cite{Goddard2004}
found that the magnetoresistance features for angles $\theta \geq 70^{\circ}$
may be attributed to Lebed, Yamaji or DCK oscillations,
depending on the azimuthal angle $\phi$.
It was also found that the Lebed and Yamaji oscillations do 
not tend to coexist at the same $\phi$~\cite{Goddard2004}. 
The Lebed oscillations dominate when the plane of rotation of 
the field is roughly perpendicular to the Q1D sheets
($\phi \approx  0^{\circ}$); the Yamaji oscillations are 
more prominent when the plane of rotation of
the field is close to that containing
the short axis of the Q2D $\alpha$ pocket
($\phi \approx  90 ^{\circ}$)~\cite{Goddard2004}. 
Applying the same analysis~\cite{Goddard2004}
to the $p=9.8$~kbar AMRO data in this paper,
the $\phi$ angles 
at which the Lebed or Yamaji oscillations dominate 
are found to be comparable 
to those at ambient pressure. 

However, an additional series of AMROs is observed 
for {\it all} $\phi$ when $\theta ~ \leq ~ 70^\circ$.
Like the Yamaji and Lebed oscillations, the extra series
is periodic in $\tan \theta$, but its
frequency is considerably higher. 
To illustrate this, Fig.~\ref{fig2}(a) shows AMRO data 
at $p=9.8$~kbar. 
Two sets of oscillations are clearly seen, both periodic in 
$\tan\theta$. The frequency of the features appearing at 
$\tan\theta~ \gtsim ~3$ show them to be the Lebed oscillations 
expected for this value of $\phi$~\cite{Goddard2004}. 
The faster oscillations are only observed at 
$\tan\theta ~ \ltsim ~3~(\theta ~ \ltsim ~ 70^{\circ})$; it is these
oscillations that we will identify below as BAMROs.
The fact that the latter oscillations are observed with a similar
frequency at all planes of rotation
suggests that they result from a rather isotropic,
Q2D quasiparticle orbit.

Given such an orbit, the $\tan \theta$
frequency of the resulting AMROs 
at a given $\phi$-angle should be proportional to 
$k_{||}^{\rm max}$, the maximum in-plane wavevector of 
the orbit
projected on to the plane of rotation of the field~\cite{House1996}. 
For oscillations arising from
an elliptical cross-section 
orbit $k_{||}^{\rm max}(\phi)$ can be fitted to the equation
\begin{equation}
k^{\rm max}_{||}(\phi) =[ k^2_a \cos^2(\phi - \zeta) +  
k^2_b \sin^2(\phi - \zeta)]^{1/2}.
\label{kmax}
\end{equation}
where $k_a$ and $k_b$ are the major and minor semi-axes 
of the ellipse respectively and 
$\zeta$ is the angle between its major axis 
and the $\phi=0^\circ$ direction~\cite{House1996}.

The $k^{\rm max}_{||} (\phi)$ values for the higher frequency
AMROs ($\tan \theta ~ \ltsim ~ 3$) at 
9.8~kbar are plotted in Fig.~\ref{fig2}(b) (red circles). 
An unconstrained fit to Eq.~\ref{kmax} implies that the 
orbit that gives rise to the oscillations is 
almost circular in cross-section with an area 
$3.8 \pm 0.1 \times 10^{19}$~m$^{-2}$. 
Within the experimental errors this value agrees with the 
area of the $\beta$-orbit determined from the SdH
frequency ($3.84 \pm 0.05 \times 10^{19}$~m$^{-2}$)
measured at 9.8~kbar.  
The good agreement strongly suggests that the\
high-frequency AMROs are BAMROs
caused by $\beta$ orbits that completely
traverse the Q1D and Q2D Fermi-surface sections;
{\it i.e.} they are only made possible 
by magnetic breakdown.

For comparison, Fig.~\ref{fig2}(b) also presents
the values of $k^{\rm max}_{||}$ 
determined from the Yamaji oscillations
due to the $\alpha$ pocket (tending to occur at
$\tan \theta ~ \gtsim ~ 3$)~\cite{Goddard2004}.
These data (blue squares) are plotted against 
$\phi$ for all planes 
of rotation at which they are observed; at the others 
the Lebed oscillations dominate~\cite{Goddard2004}. 
The dashed line is a fit to Eq.~\ref{kmax}, 
where the area is constrained by $F_\alpha$ from the SdH data
The semi-major and minor axes of the $\alpha$ pocket obtained 
in this manner are 
$2.2 \pm 0.2$ nm$^{-1}$ and $1.06 \pm 0.09$ nm$^{-1}$ 
respectively.
Therefore, in good quantitative agreement with 
earlier work~\cite{Caulfield1994,Biggs},
we find the effect of increased pressure is to make the 
$\alpha$ pocket less elongated.

\begin{figure}[t]
\centering
\includegraphics[width=8.8cm]{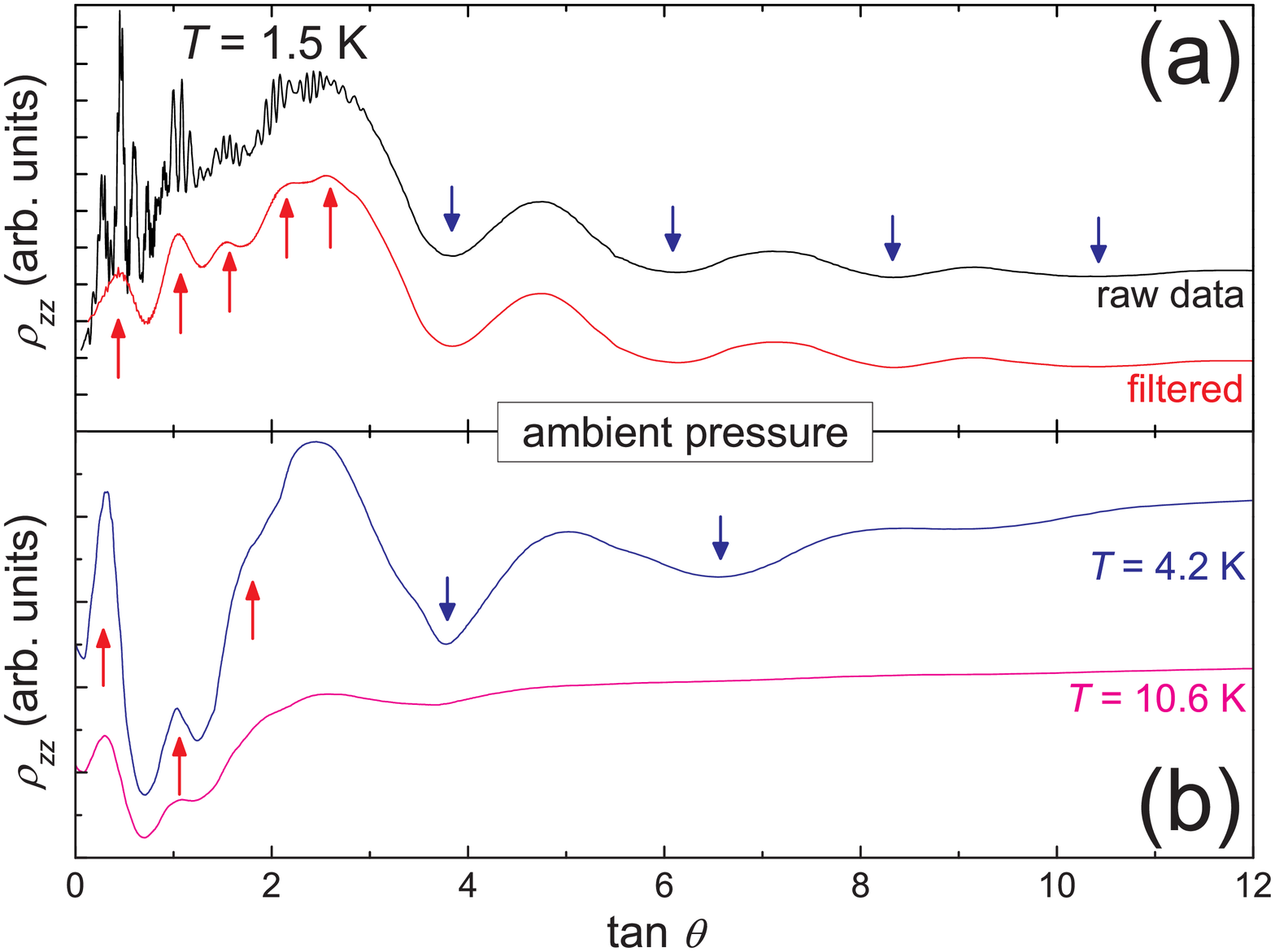}
\vspace{-5mm}
\caption{(Color online) Ambient pressure angle-dependent magnetoresistance. 
(a)~Comparison of data taken at $B=42$~T, $T=1.5$~K, 
$\phi=160^\circ$ before (upper curve) and 
after (lower curve) filtering to remove the SdH oscillations.  
(b)~Data taken at a similar $\phi$-angle with $B=45$~T. 
Two temperatures are shown, $T=4.2$~K (upper curve) 
and $T=10.6$~K. In both (a) and (b) up (red) arrows mark the 
BAMROs, down (blue) arrows the Lebed oscillations and the 
curves are offset for clarity.} 
\label{fig3}
\vspace{-5mm}
\end{figure}

Increased hydrostatic pressure 
is known to reduce the breakdown
field $B_0$ in \cuscn~\cite{Caulfield1994},
enhancing the likelihood of magnetic breakdown
(see Eq.~\ref{bd}).
Having identified BAMROs at $p=9.8$~kbar,
it is instructive to see if the same effect occurs at
ambient pressure where the breakdown probability is
lower.
Fig.~\ref{fig3}(a) shows the angle-dependent magnetoresistance 
measured at ambient pressure, $B=42$~T, $T=1.5$~K and $\phi=160^\circ$. 
The upper curve comprises raw data; as in Fig.~\ref{fig2},
Lebed oscillations are seen at $\tan\theta ~\gtsim ~3$. 
However, at lower values of $\tan\theta$ the data are 
dominated by SdH oscillations from the $\alpha$-orbit~\cite{Goddard2004}. 
The lower curve in Fig.~\ref{fig3}(a) shows the same 
data after numerical processing to remove the SdH oscillations 
(the abscissa is converted to $B\cos\theta$ and the 
data passed through a low-pass Fourier transform filter with a
100~T cut-off frequency). 
The filtering reveals the presence of AMROs, 
previously hidden by the SdH, that are periodic in 
$\tan\theta$ and almost identical to the BAMROs 
seen at 9.8~kbar. A fit to Eq.~ \ref{kmax} 
of the $\phi$-dependence of the frequency of 
these oscillations gives an orbit
area of $3.4 \pm 0.3 \times 10^{19}$~m$^{-2}$
in reasonable agreement with that obtained from
$F_\beta$ in the ambient-pressure SdH data 
($3.76 \pm 0.03 \times 10^{19}$~m$^{-2}$).

Fig.~\ref{fig3}(b) shows data taken at a similar 
$\phi$ to those in (a) but at higher $T$. 
AMROs are known to be robust at lower 
$B/T$ than magnetic quantum oscillations as they do not depend 
so strongly on thermal smearing of the Fermi 
surface~\cite{Blundell1996,Singleton2006}. 
Thus at $T=4.2$~K the SdH are no longer visible, 
whereas both the BAMROs and Lebed oscillations are 
clearly observed. Indeed, both are still 
discernible at $T=10.6$~K, albeit with a reduced 
amplitude~\cite{Singleton2006}.  
 
Thus, it appears that BAMROs are observable in \cuscn~
at ambient pressure. 
A comparison of Fig~\ref{fig2}(a), in which the BAMROs 
appear to be more prominent compared to the
background than the features seen in the filtered data of Fig~\ref{fig3}(a),
measured at the same temperature but at a higher magnetic field,
indeed suggests that the enhanced breakdown probability
at higher pressures promotes the BAMRO mechanism.
However, a more significant factor in explaining why the 
BAMROs are so clear in the high pressure data, but somewhat concealed in
ambient pressure data, is the relative strength of the SdH oscillations.
The sample used in the pressure studies exhibits 
a significantly lower Dingle
scattering time ($\tau_{\alpha} \approx 0.81$~ps)
than the sample used for the ambient-pressure 
experiments ($\tau_{\alpha} \approx 2.3$~ps).
Even though $m_{\alpha}$ decreases from $3.5m_{\rm e}$ to
$2.0m_{\rm e}$ on going from ambient pressure to 9.8~kbar (see above
and Ref.~\cite{Caulfield1994})
the Dingle scattering time is reduced by a greater factor,
greatly suppressing the SdH oscillations
in the 9.8~kbar experiments.

Elevated $T$s also suppress SdH oscillations~\cite{Singleton2006}
(Fig.~\ref{fig3}(b)), revealing the underlying BAMROs. 
The fact the BAMRO features 
survive at scattering times and $T$s at which 
the SdH cannot be observed is further evidence that their
mechanism is related to semiclassical quasiparticle trajectories
across the Fermi surface, similar to those
invoked to explain Yamaji oscillations~\cite{Goddard2004,Singleton2006}.

Therefore, we believe that, although present, BAMROs have not
previously been identified in \cuscn~ because, in general, 
angle-dependent magnetotransport measurements are performed 
at low $T$s with the cleanest possible samples~\cite{Goddard2004}. 
Under these conditions the data at the $\theta$-angles where 
BAMROs are observed are dominated by the SdH oscillations.

All AMROs are progressively damped as $\theta$ increases. 
In \cuscn,~ this is known to occur because the amplitude of the
Yamaji and Lebed oscillations is governed by the value of 
$\omega\tau$, where $\omega$ is an angular frequency
of the orbit responsible and $\tau$ is a scattering time~\cite{Singleton2006}.
The orbit frequency depends on the projection of the
magnetic field, and so $\omega \propto \cos\theta$,
leading to a decrease in $\omega \tau$ and hence AMRO
amplitude as $\theta$ increases~\cite{Singleton2006}.
However, compared to conventional AMROs,
BAMROs will have an additional damping 
factor due to Eq.~\ref{bd}, because 
in Q2D systems such as \cuscn,~$B_{0}$ 
is found to be inversely proportional to
$\cos \theta$~\cite{Shoenberg1984,Harrison1996}. 
The factor of $\cos\theta$ leads 
to additional attenuation as $\theta$ increases,
so that the BAMROs are only noticeable
for $\theta ~\ltsim ~70^\circ$.

In summary we have shown conclusive experimental evidence of BAMROs, 
angle-dependent magnetoresistance oscillations caused by magnetic breakdown.
Magnetic breakdown has been interpreted semiclassically in terms of quasiparticle 
orbits that jump gaps between Fermi surfaces in $k$-space \cite{Pippard1962}.  This 
model has been extensively explored via a detailed analysis of the magnetoresistance 
oscillations that arise in Mg due to the quantum interference of the quasiparticle 
orbits (see~\cite{Reifenberger1977} and references therein). The observation of BAMROs provides a further
compelling validation of this picture of magnetic breakdown, and in addition represents 
the only experimental manifestation of magnetic breakdown that can be described in 
purely semiclassical terms.

This work is funded by US Department of Energy
(DoE) LDRD grants 20040326ER, 20030084DR and 20070013DR, and
by EPSRC grant GR/T27341/01 (UK). NHMFL 
is supported by the NSF, DoE and the State of Florida. 
Work at Argonne was supported by
the Office of Basic Energy Sciences, US DoE 
(contract W-31-109-ENG-38). AA and AIC acknowledge support from the Royal Society, and PAG from the Oxford University Glasstone Fund.

\end{document}